\documentclass[conference, a4paper]{IEEEtran}
\IEEEoverridecommandlockouts
\usepackage{cite}
\usepackage{amsmath,amssymb,amsfonts,amsthm}
\usepackage[nolist]{acronym}
\usepackage{algorithmic}
\usepackage{hyperref}
\usepackage{scalerel}
\usepackage{graphicx}
\usepackage{textcomp}
\usepackage[table]{xcolor}
\usepackage{tikz}
\usepackage{nicefrac}
\usepackage{circuitikz}

\usepackage{booktabs}

\usepackage{multirow}

\usepackage{siunitx}

\usepackage{makecell}

\usepackage{xurl}

\newdimen\bpt
\def\mobile#1{\leavevmode 
   \bpt=#1bp \hbox to7\bpt{\kern1\bpt \lower1\bpt\vbox to12\bpt{}%
      \pdfliteral{q #1 0 0 #1 0 0 cm 1 j 2 w 0 0 5 10 re B 
         1 g 1 G  1 w .3 1.8 4.4 7 re B 
         1.5 w 2.5 .2 0 .1 re B .3 w 1.7 10 1.6 0 re B Q}%
      \hss}}

\usepackage{adjustbox}
\usetikzlibrary{calc}
\usetikzlibrary{positioning}
\usepackage{pgfplots}
\usepgfplotslibrary{groupplots}
\usepgfplotslibrary{colorbrewer}
\pgfplotsset{compat = 1.15}

\makeatletter
\let\MYcaption\@makecaption
\makeatother

\usepackage[font=footnotesize]{subcaption}

\makeatletter
\let\@makecaption\MYcaption
\makeatother

\DeclareMathOperator*{\argmax}{arg\,max}
\DeclareMathOperator*{\argmin}{arg\,min}

\usetikzlibrary{svg.path}

\def\BibTeX{{\rm B\kern-.05em{\sc i\kern-.025em b}\kern-.08em
    T\kern-.1667em\lower.7ex\hbox{E}\kern-.125emX}}

\definecolor{orcidlogocol}{HTML}{A6CE39}
\tikzset{
  orcidlogo/.pic={
    \fill[orcidlogocol] svg{M256,128c0,70.7-57.3,128-128,128C57.3,256,0,198.7,0,128C0,57.3,57.3,0,128,0C198.7,0,256,57.3,256,128z};
    \fill[white] svg{M86.3,186.2H70.9V79.1h15.4v48.4V186.2z}
                 svg{M108.9,79.1h41.6c39.6,0,57,28.3,57,53.6c0,27.5-21.5,53.6-56.8,53.6h-41.8V79.1z M124.3,172.4h24.5c34.9,0,42.9-26.5,42.9-39.7c0-21.5-13.7-39.7-43.7-39.7h-23.7V172.4z}
                 svg{M88.7,56.8c0,5.5-4.5,10.1-10.1,10.1c-5.6,0-10.1-4.6-10.1-10.1c0-5.6,4.5-10.1,10.1-10.1C84.2,46.7,88.7,51.3,88.7,56.8z};
  }
}

\newcommand\orcidicon[1]{\href{https://orcid.org/#1}{\mbox{\scalerel*{
\begin{tikzpicture}[yscale=-1,transform shape]
\pic{orcidlogo};
\end{tikzpicture}
}{|}}}}

\begin{acronym}
 \acro{R2M}{raw 2\textsuperscript{nd} moment}
 \acro{CSI}{channel state information}
 \acro{UE}{user equipment}
 \acro{EM}{electromagnetic}
 \acro{UL}{uplink}
 \acro{BS}{base station}
 \acro{TDD}{time division duplex}
 \acro{FDD}{frequency division duplex}
 \acro{ECC}{error-correcting code}
 \acro{MLD}{maximum likelihood decoding}
 \acro{HDD}{hard decision decoding}
 \acro{IF}{intermediate frequency}
 \acro{RF}{radio frequency}
 \acro{SDD}{soft decision decoding}
 \acro{NND}{neural network decoding}
 \acro{CNN}{convolutional neural network}
 \acro{ML}{maximum likelihood}
 \acro{MLE}{maximum likelihood estimation}
 \acro{GPU}{graphical processing unit}
 \acro{BP}{belief propagation}
 \acro{LTE}{Long Term Evolution}
 \acro{BER}{bit error rate}
 \acro{SNR}{signal-to-noise-ratio}
 \acro{ReLU}{rectified linear unit}
 \acro{BPSK}{binary phase shift keying}
 \acro{QPSK}{quadrature phase shift keying}
 \acro{AWGN}{additive white Gaussian noise}
 \acro{MSE}{mean squared error}
 \acro{LLR}{log-likelihood ratio}
 \acro{MAP}{maximum a posteriori}
 \acro{NVE}{normalized validation error}
 \acro{BCE}{binary cross-entropy}
 \acro{CE}{cross-entropy}
 \acro{BLER}{block error rate}
 \acro{SQR}{signal-to-quantisation-noise-ratio}
 \acro{MIMO}{multiple-input multiple-output}
 \acro{mMIMO}{massive multiple-input multiple-output}
 \acro{OFDM}{orthogonal frequency division multiplex}
 \acro{RF}{radio frequency}
 \acro{LOS}{line of sight}
 \acro{NLoS}{non-line of sight}
 \acro{NMSE}{normalized mean squared error}
 \acro{CFO}{carrier frequency offset}
 \acro{SFO}{sampling frequency offset}
 \acro{IPS}{indoor positioning system}
 \acro{TRIPS}{time-reversal IPS}
 \acro{RSSI}{received signal strength indicator}
 \acro{MIMO}{multiple-input multiple-output}
 \acro{ENoB}{effective number of bits}
 \acro{AGC}{automatic gain control}
 \acro{ADC}{analog to digital converter}
 \acro{ADCs}{analog to digital converters}
 \acro{FB}{front bandpass}
 \acro{FPGA}{field programmable gate array}
 \acro{JSDM}{Joint Spatial Division and Multiplexing}
 \acro{NN}{neural network}
 \acro{IF}{intermediate frequency}
 \acro{LoS}{line-of-sight}
 \acro{NLoS}{non-line-of-sight}
 \acro{DSP}{digital signal processing}
 \acro{AFE}{analog front end}
 \acro{SQNR}{signal-to-quantisation-noise-ratio}
 \acro{SINR}{signal-to-interference-noise-ratio}
 \acro{ENoB}{effective number of bits}
 \acro{PCB}{printed circuit board}
 \acro{EVM}{error vector mangnitude}
 \acro{CDF}{cumulative distribution function}
 \acro{MRC}{maximum ratio combining}
 \acro{MRP}{maximum ratio precoding}
 \acro{MRT}{maximum ratio transmission}
 \acro{DeepL}{deep-learning}
 \acro{DL}{downlink}
 \acro{SISO}{single-input single-output}
 \acro{SGD}{stochastic gradient descent}
 \acro{CP}{cyclic prefix}
 \acro{MISO}{Multiple Input Single Output}
 \acro{LMMSE}{linear minimum mean square error}
 \acro{ZF}{zero forcing}
 \acro{USRP}{universal software radio peripheral}
 \acro{RNN}{recurrent neural network}
 \acro{GRU}{gated recurrent unit}
 \acro{LSTM}{long short-term memory}
 \acro{NTM}{neural turing machine}
 \acro{DNC}{differentiable neural computer}
 \acro{TCN}{temporal convolutional network}
 \acro{FCL}{fully connected layer}
 \acro{MANN}{memory augmented neural network}
 \acro{RNN}{recurrent neural network}
 \acro{DNN}{deep neural network}
 \acro{FIR}{finite impulse response}
 \acro{BPTT}{back-propagation through time}
 \acro{GAN}{generative adversarial network}
 \acro{ELU}{exponential linear unit}
 \acro{tanh}{hyperbolic tangent}
 \acro{BICM}{bit-interleaved coded modulation}
 \acro{OTA}{over-the-air}
 \acro{IM}{intensity modulation}
 \acro{DD}{direct detection}
 \acro{RL}{reinforcement learning}
 \acro{SDR}{software-defined radio}
 \acro{WGAN}{Wasserstein generative adversarial network}
 \acro{BMD}{bit-metric decoding}
 \acro{BMI}{bit-wise mutual information}
 \acro{LDPC}{low-density parity-check}
 \acro{IDD}{iterative demapping and decoding}
 \acro{JSD}{Jensen-Shannon divergence}
 \acro{MMSE}{minimum mean square error}
 \acro{FFT}{fast Fourier transform}
 \acro{DFT}{discrete Fourier transform}
 \acro{IFFT}{inverse fast Fourier transform}
 \acro{QAM}{quadrature amplitude modulation}
 \acro{EMD}{earth mover's distance}
 \acro{TDL}{tapped delay line}
 \acro{KL}{Kullback-Leibler}
 \acro{PRACH}{physical random access channel}
 \acro{URLLC}{ultra-reliable low-latency communication}
 \acro{ANOMA}{asynchronous non-orthogonal multiple access}
 \acro{FEC}{forward error correction}
 \acro{PAPR}{peak-to-average power ratio}
 \acro{APP}{a posteriori probability}
 \acro{COTS}{commercial off-the-shelf}
 \acro{PLL}{phase locked loop}
 \acro{STO}{sampling time offset}
 \acro{SFO}{sampling frequency offset}
 \acro{CFO}{carrier frequency offset}
 \acro{CPO}{carrier phase offset}
 \acro{CSI}{channel state information}
 \acro{GNSS}{global navigation satellite system}
 \acro{ELAA}{extremely large aperture array}
 \acro{UE}{user equipment}
 \acro{DICHASUS}{\underline{Di}stributed \underline{Cha}nnel \underline{S}ounder by \underline{U}niversity of \underline{S}tuttgart}
 \acro{JCaS}{Joint Communication and Sensing}
 \acro{CT}{Continuity}
 \acro{TW}{Trustworthiness}
 \acro{KS}{Kruskal's stress}
 \acro{PCA}{principal component analysis}
 \acro{AoA}{angle of arrival}
 \acro{ToA}{time of arrival}
 \acro{RD}{Rajski's distance}
 \acro{ADP}{angle-delay profile}
 \acro{MPC}{multipath component}
 \acro{CIR}{channel impulse response}
 \acro{MAE}{mean absolute error}
 \acro{MDS}{multidimensional scaling}
 \acro{t-SNE}{t-distributed stochastic neighbor embedding}
 \acro{SM}{Sammon's mapping}
 \acro{CIRA}{channel impulse response amplitude}
 \acro{CS}{cosine similarity}
 \acro{FCF}{forward charting function}
 \acro{CMD}{correlation matrix distance}
 \acro{6G}{Sixth generation}
 \acro{JEPA}{joint-embedding predictive architecture}
 \acro{MAC}{medium access control}
 \acro{WSS}{wide-sense stationary}
 \acro{UPA}{uniform planar array}
 \acro{UAV}{unmanned aerial vehicle}
\end{acronym}

\definecolor{mittelblau}{RGB}{0, 126, 198}
\definecolor{violettblau}{cmyk}{0.9, 0.6, 0, 0}
\definecolor{rot}{RGB}{238, 28 35}
\definecolor{apfelgruen}{RGB}{140, 198, 62}
\definecolor{gelb}{RGB}{1, 221, 0}
\definecolor{orange}{RGB}{244, 111, 33}
\definecolor{pink}{RGB}{237, 0, 140}
\definecolor{lila}{RGB}{128, 10, 145}
\definecolor{hellgrau}{RGB}{224, 224, 224}
\definecolor{mittelgrau}{RGB}{128, 128, 128}
\definecolor{dunkelgrau}{RGB}{80,80,80}
\definecolor{anthrazit}{RGB}{19, 31, 31}

\begin{document}

\title{Three-Dimensional Radio Localization: A Channel Charting-Based Approach\\
\thanks{This work is supported by the German Federal Ministry of Research, Technology and Space (BMFTR) within the projects Open6GHub (grant no. 16KISK019) and KOMSENS-6G (grant no. 16KISK113).}
}

\author{\IEEEauthorblockN{Phillip Stephan\textsuperscript{\orcidicon{0009-0007-4036-668X}}, Florian Euchner\textsuperscript{\orcidicon{0000-0002-8090-1188}}, Stephan ten Brink\textsuperscript{\orcidicon{0000-0003-1502-2571}} \\}


\IEEEauthorblockA{
Institute of Telecommunications, Pfaffenwaldring 47, University of  Stuttgart, 70569 Stuttgart, Germany \\ \{stephan,euchner,tenbrink\}@inue.uni-stuttgart.de
}
}

\maketitle

\begin{abstract}
Channel charting creates a low-dimensional representation of the radio environment in a self-supervised manner using manifold learning.
Preserving relative spatial distances in the latent space, channel charting is well suited to support user localization.
While prior work on channel charting has mainly focused on two-dimensional scenarios, real-world environments are inherently three-dimensional.
In this work, we investigate two distinct three-dimensional indoor localization scenarios using simulated, but realistic ray tracing-based datasets: a factory hall with a three-dimensional spatial distribution of datapoints, and a multistory building where each floor exhibits a two-dimensional datapoint distribution.
For the first scenario, we apply the concept of augmented channel charting, which combines classical localization and channel charting, to a three-dimensional setting.
For the second scenario, we introduce multistory channel charting, a two-stage approach consisting of floor classification via clustering followed by the training of a dedicated expert neural network for channel charting on each individual floor, thereby enhancing the channel charting performance.
In addition, we propose a novel feature engineering method designed to extract sparse features from the beamspace channel state information that are suitable for localization.
\end{abstract}


\section{Introduction}\label{sec:intro}
The precise localization of mobile devices enables various applications, such as navigation services, and plays an important role in the modern world.
While navigation systems based on \acp{GNSS} serve as the standard solution for outdoor positioning, their performance degrades significantly in scenarios with obstructed signals, such as indoor environments and dense urban street canyons.
Therefore, exploiting the existing wireless communication infrastructure for localization represents a viable alternative and has become a prominent field of research \cite{wen_5g_localization_survey}.
Conventional model-based localization techniques typically rely on geometric triangulation or trilateration using features such as \ac{RSSI}, \ac{ToA}, and \ac{AoA} measured at the \ac{BS} to estimate user positions.
However, the chaotic nature of radio wave propagation, particularly under \ac{NLoS} conditions, makes such model-based estimators prone to errors.
In contrast, data-driven methods based on \acp{DNN} for \ac{CSI} fingerprinting have shown to achieve high localization accuracy even in complex propagation environments \cite{savic2015fingerprinting, vieira2017deep, cc_features_ferrand}.
Despite these advances, the practical deployment of \ac{CSI} fingerprinting is still limited by the substantial effort required to acquire labeled ground truth position data for training.

Channel charting \cite{studer_cc}, on the other hand, employs manifold learning techniques to create a low-dimensional representation of the radio environment in a self-supervised manner.
Using similarity relationships within the \ac{CSI}, the resulting channel chart ideally preserves the global structure of the underlying physical environment.
The channel chart may be subject to scaling and/or rotation with respect to the physical coordinate axes, which is inconsequential for many applications, including handover prediction \cite{kazemi_cc_snr_prediction}, beam prediction \cite{yassine_beam_prediction}, and channel prediction \cite{wcnc2025}.
Previous papers have also demonstrated that channel charting can be used to enhance absolute user localization \cite{pihlajasalo2020absolute, taner_cc_real_world_coordinates, asilomar2023}.
While prior work on channel charting has mainly addressed two-dimensional environments, the physical world is inherently three-dimensional, and estimating the height of \acp{UE} can be crucial for various applications, such as locating \acp{UAV} or \acp{UE} in a multistory building.
In \cite{karmanov2021wiclusterpassiveindoor2d3d}, the authors proposed a weakly supervised approach to the three-dimensional indoor localization of passive objects that combines channel charting methods with a specific zone loss, which requires knowledge of the floor plan of the building.

\begin{table}
    \caption{Symbols and notations used in this paper}\label{tab:notations}
    \vspace{-0.18cm}
    \centering
    \begin{tabular}{r | l}
        $\mathbf A$, $\mathbf b$ & \begin{tabular}{@{}l@{}}Bold letters: Uppercase for matrices and tensors (here \vspace{-0.07cm}\\ as multidimensional arrays), lowercase for vectors\end{tabular}\\
        $m, N$ & Italic uppercase or lowercase letters: Scalars\\
        $\mathbf{A}^{(l)}$ & Superscript letters: indexing time instant $l$ of tensor $\mathbf{A}$\\
        $\mathbf{A}_{ijk}$ & \begin{tabular}{@{}l@{}}Subscript letters: indexing elements \vspace{-0.07cm}\\ along axes $i,j,k$ of tensor $\mathbf{A}$\end{tabular}\\
        \begin{tabular}{l@{}l@{}l@{}}$\mathbf A_{i::}$ \vspace{-0.07cm}\\$\mathbf A_{ij:}$ \vspace{-0.07cm}\\$\mathbf A_{i:k}$ \end{tabular} & \begin{tabular}{@{}l@{}l@{}}Sub-matrix (and sub-vector) of elements in $i$\textsuperscript{th} entry \vspace{-0.07cm}\\of the first dim. (and $j$\textsuperscript{th} entry of the second dim. \vspace{-0.07cm}\\or $k$\textsuperscript{th} entry of the third dim.) of tensor $\mathbf A$\end{tabular}\\
        $\lVert \mathbf b \rVert$ & Euclidean norm of vector $\mathbf b$ \vspace{0.02cm}\\
        $\mathbf A^\mathrm{H}$, $\mathbf b^\mathrm{H}$ & Conjugate transpose of matrix $\mathbf A$ (or vector $\mathbf b$)\\
    \end{tabular}
    \vspace{-0.3cm}
\end{table}

\begin{figure*}
    \centering
    \begin{subfigure}[b]{0.22\textwidth}
        \centering
        \includegraphics[width=\textwidth, clip]{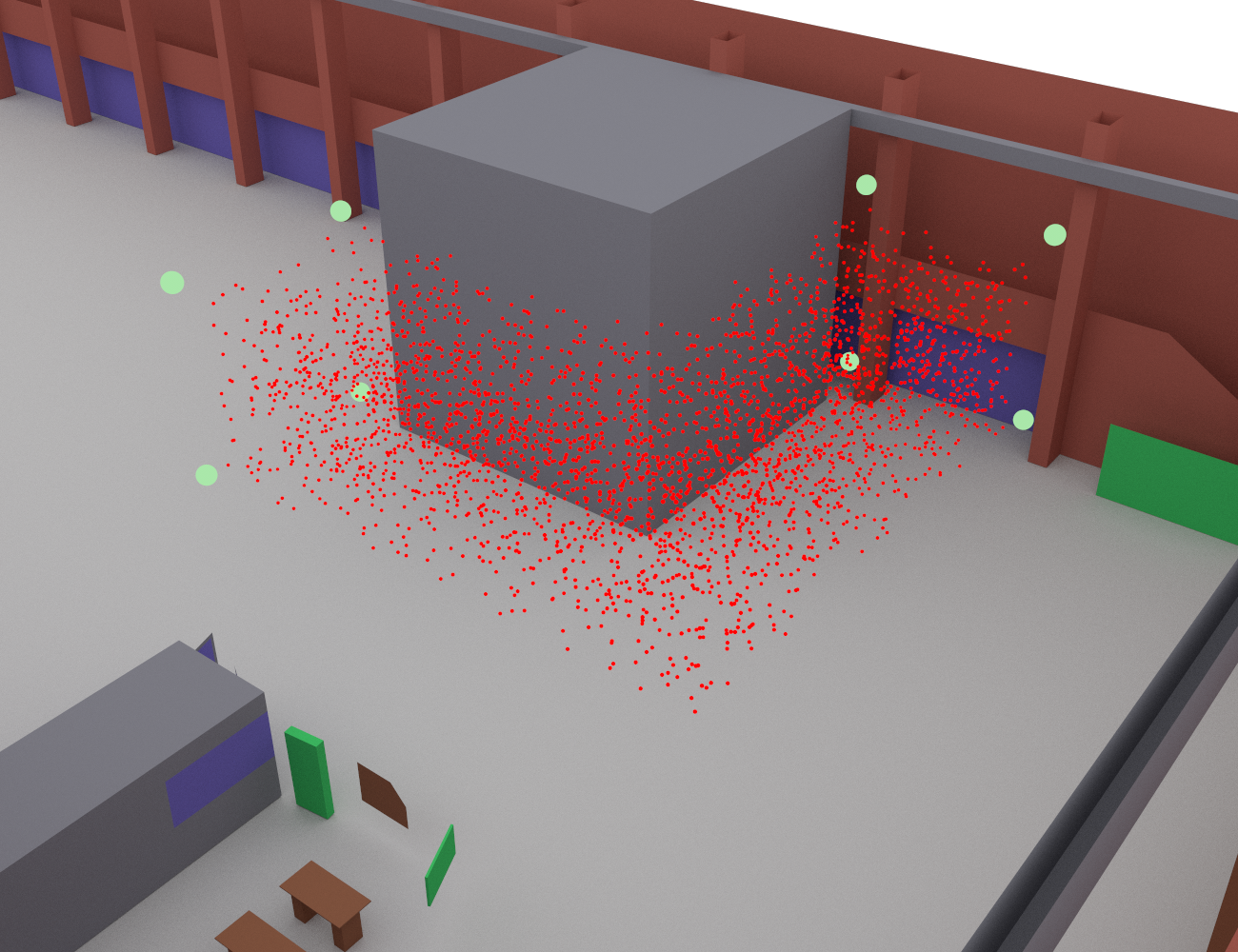}
        \caption{Factory hall: Scene}
        \label{fig:dataset_scenario1_scene}
    \end{subfigure}
    \begin{subfigure}[b]{0.27\textwidth}
        \centering
        \includegraphics[width=0.95\textwidth, clip]{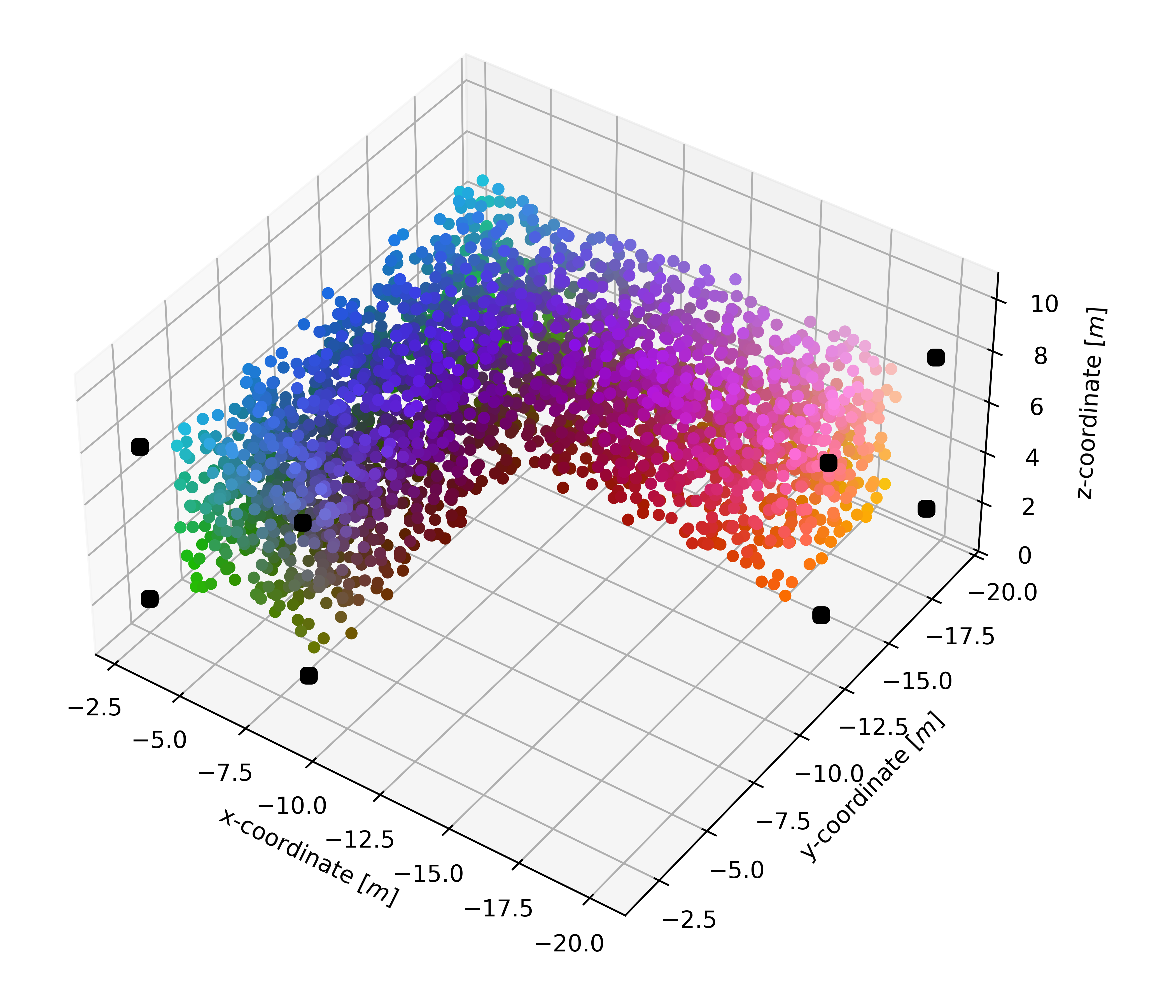}
        \caption{Factory hall: \ac{UE} positions}
        \label{fig:dataset_scenario1_positions}
    \end{subfigure}
    \begin{subfigure}[b]{0.22\textwidth}
        \centering
        \includegraphics[width=\textwidth, clip]{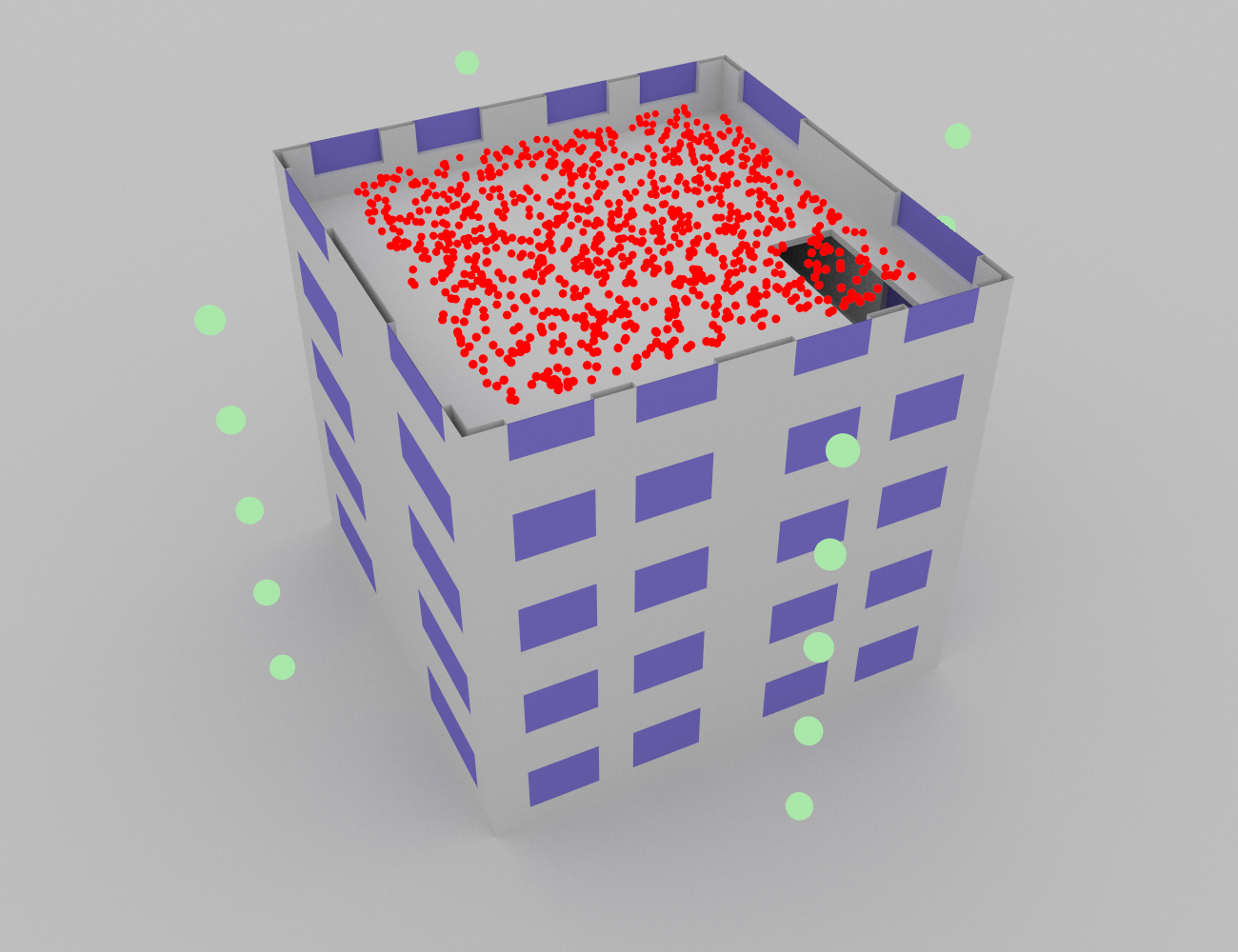}
        \caption{Multistory building: Scene}
        \label{fig:dataset_scenario2_scene}
    \end{subfigure}
    \begin{subfigure}[b]{0.27\textwidth}
        \centering
        \includegraphics[width=0.95\textwidth, clip]{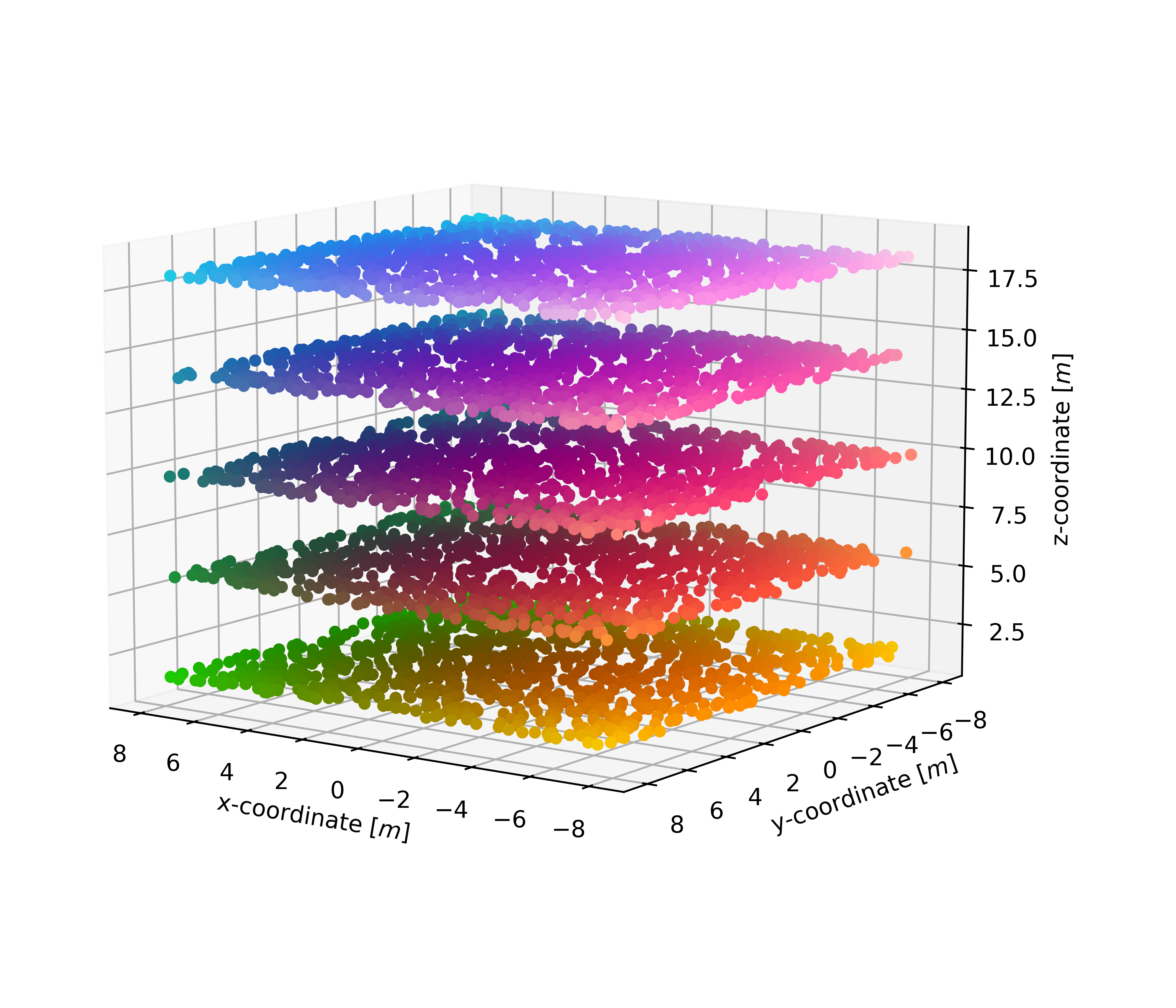}
        \caption{Multistory building: \ac{UE} positions}
        \label{fig:dataset_scenario2_positions}
    \end{subfigure}
    \vspace{-0.5cm}
    \caption{Visualization of the ray tracing-based datasets: The figure shows (a) an image of the synthetic factory hall (scenario 1), and (c) an image of the synthetic multistory building (scenario 2), both with the \ac{BS} antenna arrays marked as green dots and the ``ground truth'' \ac{UE} positions as red dots, and (b) and (d) the respective scatter plots of the ``ground truth'' positions, each with a unique color gradient. (b) additionally marks the array positions as black dots.}
    \label{fig:dataset}
\end{figure*}

\subsection{Contributions}
In contrast to \cite{karmanov2021wiclusterpassiveindoor2d3d}, we introduce a self-supervised framework for the three-dimensional radio localization of active \acp{UE} based on channel charting that does not require any floor plan of the building.
We investigate two three-dimensional indoor localization scenarios using realistic ray tracing-based datasets: a factory hall with a three-dimensional spatial distribution of datapoints, and a multistory building where each floor exhibits a two-dimensional datapoint distribution.
For the first scenario, we apply the concept of augmented channel charting from our previous work \cite{asilomar2023}, which combines classical localization and channel charting, to a three-dimensional setting.
For the second scenario, we introduce multistory channel charting, a two-stage approach consisting of floor classification via clustering followed by the training of a dedicated expert neural network for channel charting on each individual floor, thereby enhancing the channel charting performance.
Furthermore, we introduce a new feature engineering method based on beamspace \ac{CSI}, particularly suited for massive MIMO systems deploying \acp{UPA}\footnote{The ray tracing and channel charting source code used in this work are publicly available at \url{https://github.com/phillipstephan/3D-ChannelCharting}}.

\subsection{Outline}
The remainder of this paper is organized as follows.
Section~\ref{sec:dataset} presents two distinct localization scenarios, each associated with its own ray tracing-based \ac{CSI} dataset.
Section~\ref{sec:classical_triangulation} briefly reviews the classical localization approach relying on \ac{AoA} estimation and triangulation.
Subsequently, Section~\ref{sec:channel_charting} outlines the conventional channel charting framework and the concept of augmented channel charting, and introduces a new feature engineering method, as well as the concept of multistory channel charting.
The localization performance of the applied methods is then evaluated in Section~\ref{sec:results}.
Finally, Section~\ref{sec:conclusion} provides a summary and discusses possible future research activities.
The symbols and notations used in this paper are listed in Table~\ref{tab:notations}.

\subsection{Limitations}
The principal limitation of this work is the use of ray tracing-based \ac{CSI} for evaluation.
Future research may apply the proposed concepts on real channel measurements in order to obtain more representative and practically relevant results.
Moreover, the antenna configurations and propagation environments considered in both scenarios are fortunate for localization. Subsequent work should therefore address this limitation by validating the proposed method under alternative antenna configurations and diverse radio environments.

\section{System Model and Datasets}\label{sec:dataset}
We consider a wireless communication system comprising a single-antenna \ac{UE}, and a \ac{BS} equipped with $B$ distributed \acp{UPA} located at known positions $\mathbf{p}_b \in \mathbb{R}^3$.
The known orientation matrix $\boldsymbol{\Omega}_b \in \mathbb{R}^{3 \times 3}$ of array $b$ is defined by the orthogonal unit vectors of its local coordinate system with respect to the global coordinate system.
Each array consists of $M_\mathrm{row} \times M_\mathrm{col}$ antenna elements, which are assumed to be synchronized in frequency across all antennas, and additionally in time and phase within each array.
The system operates at a carrier frequency of $3.438\,\mathrm{GHz}$ and employs $N_\mathrm{sub} = 64$ \ac{OFDM} subcarriers over a bandwidth of $50\,\mathrm{MHz}$.
At each time instant $l$, the \ac{CSI} between the \ac{UE} and all individual \ac{BS} antennas is acquired for all subcarriers and stored as a tensor $\mathbf H^{(l)} \in \mathbb C^{B \times M_\mathrm{row} \times M_\mathrm{col} \times N_\mathrm{sub}}$.
The \ac{CSI} tensor is stored with the corresponding ``ground truth'' \ac{UE} position $\mathbf x^{(l)}$ in the dataset
\[
    \text{Dataset}: \mathcal D = \left\{ \left(\mathbf H^{(l)}, \mathbf x^{(l)} \right) \right\}_{l = 1, \ldots, L}.
\]
In the remainder of this section, we describe two distinct localization scenarios.
For each scenario, a separate \ac{CSI} dataset is generated using Sionna Ray Tracing \cite{sionna}.

\subsection{Scenario 1: Factory Hall}
The first scenario is a realistic indoor environment based on a point cloud measured in an actual factory hall \cite{dataset-dichasus-adxx}.
Fig.~\ref{fig:dataset_scenario1_scene} shows the environment, where the array locations are marked in green and the \ac{UE} positions are marked in red.
The \ac{BS} deploys $B = 8$ distributed arrays, each comprising $M_\mathrm{row} \times M_\mathrm{col} = 8 \times 8$ antenna elements.
A large metallic cube in the center of the environment obstructs the \ac{LoS} paths for a subset of the arrays over a significant portion of the area.
The \ac{UE} collects \ac{CSI} at positions distributed around this metal cube.
These positions are illustrated in Fig.~\ref{fig:dataset_scenario1_positions} using a unique color gradient for subsequent evaluation of the localization methods, along with the array positions, which are indicated in black.
The \ac{UE} positions span a horizontal area of $16\, \si{\meter} \times 16\, \si{\meter}$ and vertical heights ranging from $2\, \si{\meter}$ to $8\,\si{\meter}$.
The dataset corresponding to scenario 1 is denoted by $\mathcal{D}_\mathrm{factory}$ and contains $L_\mathrm{factory} = 5000$ datapoints.

\subsection{Scenario 2: Multistory Building}
The second scenario considers a simplified, hypothetical multistory office building comprising five floors.
The \ac{BS} deploys $B = 5 \times 4$ distributed arrays, each equipped with $M_\mathrm{row} \times M_\mathrm{col} = 2 \times 4$ antenna elements.
As visualized in Fig.~\ref{fig:dataset_scenario2_scene}, the \ac{BS} arrays, indicated by the green dots, are located outside the building, e.g., along the facades of adjacent buildings.
This scenario is intentionally designed to demonstrate that a \ac{BS} located outside the building can still perform channel charting.
On each floor, the \ac{UE} positions, indicated as red dots, are distributed over a two-dimensional area of $16\,\si{\meter} \times 16\,\si{\meter}$.
The corresponding three-dimensional locations (including the floor-dependent height ranging from $1.50\,\si{\meter}$ to $18.50\,\si{\meter}$) are also visualized with color gradient in Fig.~\ref{fig:dataset_scenario2_positions}.
The dataset associated with scenario 2 is denoted by $\mathcal{D}_\mathrm{multi}$ and contains $L_\mathrm{multi} = 5000$ datapoints ($1000$ samples per floor).

\section{Classical Localization: AoA + Triangulation}\label{sec:classical_triangulation}
We implement a classical localization method based on \ac{AoA} estimation and triangulation.
The system's lag of time synchronization prevents the incorporation of \ac{ToA}-based multilateration as in \cite{asilomar2023}.
To estimate the \acp{AoA}, we first compute the azimuth and elevation covariance matrices for each array $b$ and at each time instant $l$ as
\vspace{0.05cm}%
\[
    \mathbf R_{\mathrm{az},b}^{(l)} = \sum_{m_\mathrm{r} = 1}^{M_\mathrm{r}} \sum_{n = 1}^{N_\mathrm{sub}} \left(\mathbf H_{bm_\mathrm{r}:n}^{(l)}\right) \left(\mathbf H_{bm_\mathrm{r}:n}^{(l)}\right)^\mathrm{H},
\]
\vspace{0.01cm}%
and
\vspace{0.02cm}%
\[
    \mathbf R_{\mathrm{el},b}^{(l)} = \sum_{m_\mathrm{c} = 1}^{M_\mathrm{c}} \sum_{n = 1}^{N_\mathrm{sub}} \left(\mathbf H_{b:m_\mathrm{c}n}^{(l)}\right) \left(\mathbf H_{b:m_\mathrm{c}n}^{(l)}\right)^\mathrm{H},
\]
\vspace{0.05cm}%
respectively.
These matrices are used by the root-MUSIC algorithm to derive the estimated azimuth \acp{AoA} $\hat{\alpha}_{\mathrm{az},b}^{(l)}$ and elevation \acp{AoA} $\hat{\alpha}_{\mathrm{el},b}^{(l)}$.
Subsequently, we perform three-dimensional triangulation similar to \cite{henninger_3D_localization}.
As is common, the \ac{AoA} likelihood function can be described using the von Mises-Fisher distribution as:
\vspace{0.06cm}%
\[
    \mathcal L_\mathrm{AoA}(\mathbf x) = \prod_{b = 1}^B \frac{\kappa_b}{4\pi \sinh{\kappa_b}} \exp\left( \mathbf{\hat{u}}_b^T \boldsymbol{\Omega}_b^T \frac{\mathbf{x}-\mathbf{p}_b}{\lVert\mathbf{x}-\mathbf{p}_b\rVert}\right),
    \label{eq:aoalikelihood}
\]
\vspace{0.12cm}%
where $\mathbf{\hat{u}}_b$ represents the unit direction vector derived from the estimated angles $\hat \alpha_{\mathrm{az},b}$ and $\hat \alpha_{\mathrm{el},b}$ in the local coordinate system of array $b$ at position $\mathbf{p}_b$ with transformation matrix $\boldsymbol{\Omega}_b$.
The parameter $\kappa_b$ is a concentration parameter derived from a heuristic related to the delay spread at array $b$.
Finally, the position estimate $\mathbf { \hat x }^{(l)}$ is obtained by \ac{MLE}:
\vspace{0.11cm}%
\[
    \mathbf { \hat x }^{(l)} = \argmax_{\mathbf x} \mathcal L^{(l)}_\mathrm{AoA}(\mathbf x).
\]

\section{Channel Charting}\label{sec:channel_charting}

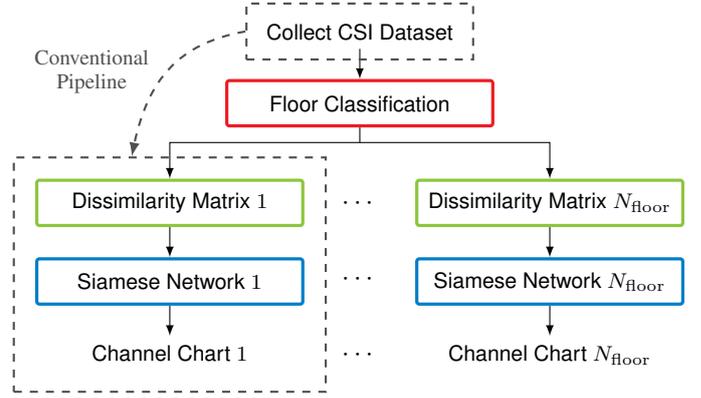
\begin{figure}
    \centering
    \usetikzlibrary{positioning, shapes, calc, fit, arrows.meta}

\begin{tikzpicture}
    
    \node (dataset) [font={\footnotesize\sffamily}] {Collect CSI Dataset};

    \node (class) [rectangle, below = 0.4cm of dataset, draw = rot, very thick, rounded corners = 1pt, inner sep = 4pt, align = center, font={\footnotesize\sffamily}, minimum width = 3.5cm, minimum height = 0.6cm] {Floor Classification};

    \coordinate (dm_center) at ([yshift=-0.2cm]class.south); 

    \node (dm1) [rectangle, minimum width = 3.5cm, minimum height = 0.6cm, draw = apfelgruen, very thick, rounded corners = 1pt, inner sep = 4pt, align = center, font={\footnotesize\sffamily}] at ([xshift=-2.5cm,yshift=-0.8cm]dm_center) {Dissimilarity Matrix $1$};
    
    \node (dr1) [rectangle, below = 0.4cm of dm1, minimum width = 3.5cm, minimum height = 0.6cm, draw = mittelblau, very thick, rounded corners = 1pt, inner sep = 4pt, align = center, font={\footnotesize\sffamily}] {Siamese Network $1$};
    
    \node (cc1) [below = 0.4cm of dr1, font={\footnotesize\sffamily}] {Channel Chart $1$};
    
    \node (dots) [below = 0.6cm of dm_center] {$\cdots$};
    \node (dots2) [below = 0.6cm of dots] {$\cdots$};
    \node (dots3) [below = 0.6cm of dots2] {$\cdots$};

    \node (dmN) [rectangle, minimum width = 3.5cm, minimum height = 0.6cm, draw = apfelgruen, very thick, rounded corners = 1pt, inner sep = 4pt, align = center, font={\footnotesize\sffamily}] at ([xshift=2.5cm,yshift=-0.8cm]dm_center) {Dissimilarity Matrix $N_\mathrm{floor}$};
    
    \node (drN) [rectangle, below = 0.4cm of dmN, minimum width = 3.5cm, minimum height = 0.6cm, draw = mittelblau, very thick, rounded corners = 1pt, inner sep = 4pt, align = center, font={\footnotesize\sffamily}] {Siamese Network $N_\mathrm{floor}$};
    
    \node (ccN) [below = 0.4cm of drN, font={\footnotesize\sffamily}] {Channel Chart $N_\mathrm{floor}$};

    \draw [-latex] (dataset) -- (class);
    \draw (class.south) -- (dm_center);
    \draw [-latex] (dm_center) -| (dm1.north);
    \draw [-latex] (dm_center) -| (dmN.north);
    \draw [-latex] (dm1) -- (dr1);
    \draw [-latex] (dr1) -- (cc1);
    \draw [-latex] (dmN) -- (drN);
    \draw [-latex] (drN) -- (ccN);

    
    \node (box_branch) [fit=(dm1) (cc1), thick, draw=dunkelgrau, dashed, inner sep=8pt] {};
    
    \node (box_data) [fit=(dataset), thick, draw=dunkelgrau, dashed, inner sep=4pt] {};

    \draw[dashed, -latex, thick, dunkelgrau] 
            (box_data.west) 
            to[out=190, in=80, looseness=1.1] 
            node[midway, left, font=\footnotesize, align=center, xshift=-0.1cm] {Conventional\\Pipeline} 
            ([xshift=-0.5cm]box_branch.north);

\end{tikzpicture}
    \vspace{-0.2cm}
    \caption{Conventional and multistory channel charting pipeline.}
    \label{fig:cc_pipeline_multistory}
\end{figure}

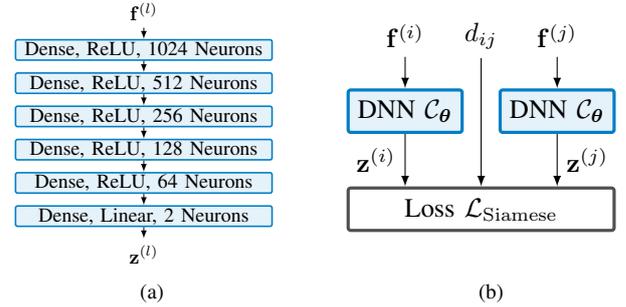
\begin{figure}
    \centering
    \begin{subfigure}[b]{0.49\columnwidth}
        \centering
        \scalebox{0.75}{
        \begin{tikzpicture}

\tikzstyle{box1} = [rectangle, draw = mittelblau, thick, rounded corners=1pt,inner sep = 1pt, align = center, fill = mittelblau!10!white,minimum width=4.5cm]

    \node (input) [anchor = south] {$\mathbf f^{(l)}$};
	\node (l1) [box1, below = 0.2cm of input] {Dense, ReLU, 1024 Neurons};
	\node (l2) [box1, below = 0.2cm of l1] {Dense, ReLU, 512 Neurons};
	\node (l3) [box1, below = 0.2cm of l2] {Dense, ReLU, 256 Neurons};
	\node (l4) [box1, below = 0.2cm of l3] {Dense, ReLU, 128 Neurons};
	\node (l5) [box1, below = 0.2cm of l4] {Dense, ReLU, 64 Neurons};
	\node (l6) [box1, below = 0.2cm of l5] {Dense, Linear, 2 Neurons};

	\node (output) [anchor = north] at ($(l6.south) + (0, -0.2)$) {$\mathbf z^{(l)}$};

    \draw [-latex] (input) -- (l1);
    \draw [-latex] (l1) -- (l2);
    \draw [-latex] (l2) -- (l3);
    \draw [-latex] (l3) -- (l4);
    \draw [-latex] (l4) -- (l5);
	\draw [-latex] (l5) -- (l6);
	\draw [-latex] (l6) -- (output);
\end{tikzpicture}
        }
        \vspace{-0.05cm}
        \caption{}
        \label{fig:dnn_structure}
    \end{subfigure}
    \begin{subfigure}[b]{0.49\columnwidth}
        \centering
        \scalebox{1.0}{
\begin{tikzpicture}
    
    \tikzstyle{box1} = [rectangle, draw = mittelblau, very thick, rounded corners=1pt,inner sep = 3pt, align = center, fill = mittelblau!10!white, minimum height=16pt,minimum width=1.5cm]
    \tikzstyle{box2} = [rectangle, draw = dunkelgrau, very thick, rounded corners=1pt,inner sep = 3pt, align = center, minimum height=16pt,minimum width=3.5cm]

    \node (in_1) at (2,0){$\mathbf{f}^{(i)}$};
    \node (in_2) at (4,0){$\mathbf{f}^{(j)}$};
    
    \node (in_3) at (3,0){$d_{ij}$};
    
    \node[box1] (dnn_1) at (2,-1.0) {DNN $\mathcal{C}_{\boldsymbol{\theta}}$};
    \node[box1] (dnn_2) at (4,-1.0) {DNN $\mathcal{C}_{\boldsymbol{\theta}}$};
    
    
    \node[box2] (contrastive_loss) at (3,-2.3) {Loss $\mathcal{L}_\mathrm{Siamese}$};
    
    \draw [-latex]  (in_1.south) -- (dnn_1.north)
    node[midway,anchor=east]{};
    \draw [-latex]  (in_2.south) -- (dnn_2.north)
    node[midway,anchor=east]{};
    
    \draw [-latex]  (in_3.south) -- (contrastive_loss)
    node[midway,anchor=east]{};
    
    
    \draw [-latex]  (dnn_1.south) -- (dnn_1.south|-contrastive_loss.north)
    node[midway,anchor=east]{$\mathbf{z}^{(i)}$};
    \draw [-latex]  (dnn_2.south) -- (dnn_2.south|-contrastive_loss.north)
    node[midway,anchor=west]{$\mathbf{z}^{(j)}$};
\end{tikzpicture}
        }
        \vspace{0.45cm}
        \caption{}
        \label{fig:siamese_structure}
    \end{subfigure}
    \vspace{-0.6cm}
    \caption{Neural network structures: (a) DNN as forward charting function, and (b) Siamese network for the training process of channel charting.}
    \label{fig:neural_networks}
\end{figure}

\subsection{Conventional Channel Charting}
As a conventional method, we apply \emph{dissimilarity metric-based channel charting} similar to \cite{stephan2024angle} in a three-dimensional environment.
The conventional pipeline, depicted in Fig.~\ref{fig:cc_pipeline_multistory}, consists of two main steps, namely the computation of a dissimilarity matrix from the measured \ac{CSI}, followed by dimensionality reduction using a Siamese neural network to learn the channel chart.
In the first step, pairwise dissimilarities (``pseudo-distances'') $d_{ij}$ between all pairs of datapoints with indices $i$ and $j$ in the dataset are computed using the \emph{geodesic angle delay profile} dissimilarity metric from \cite{stephan2024angle}.
This metric evaluates the squared cosine similarity at each antenna array and each time-domain tap within a short time window that captures the majority of the received signal power.
To obtain globally meaningful dissimilarities, we apply a shortest path algorithm.
In the second step, we learn the \acf{FCF} $\mathbf {z}^{(l)} = \mathcal C_{\boldsymbol{\theta}}(\mathbf f^{(l)})$, which maps the \ac{CSI} feature vector $\mathbf{f}^{(l)}$ to its three-dimensional channel chart position $\mathbf{z}^{(l)} \in \mathbb{R}^3$.
The \ac{FCF} is realized as a \ac{DNN} with trainable parameters $\boldsymbol{\theta}$, which facilitates inference for previously unseen datapoints.
The architecture of all \acp{DNN} employed throughout this work is illustrated in Fig.~\ref{fig:dnn_structure}.
The \ac{CSI} feature vectors are computed in a dedicated feature engineering step (see Section~\ref{sec:beamspace_features}) from the \ac{CSI}, with the objective of preserving meaningful information while omitting redundancies.
The Siamese network structure, illustrated in Fig.~\ref{fig:siamese_structure}, enables the \ac{DNN} to jointly process two input feature vectors $\mathbf{f}^{(i)}$ and $\mathbf{f}^{(j)}$ during training.
The network produces the corresponding channel chart positions $\mathbf{z}^{(i)}$ and $\mathbf{z}^{(j)}$, which are optimized such that their Euclidean distance matches the respective dissimilarity $d_{ij}$.
This is achieved by the Siamese loss
\[
\mathcal{L}_\mathrm{Siamese}=\sum\nolimits_{i=1}^{L-1}\sum\nolimits_{j=i+1}^L \frac{\left(d_{ij}-\Vert\mathbf{z}^{(i)}-\mathbf{z}^{(j)}\Vert\right)^2}{d_{ij} + \beta},
\label{eq:siameseloss}
\]
where $L$ denotes the number of training samples and the hyperparameter $\beta$ controls the weighting of absolute and normalized squared error.
Note that the resulting channel chart positions $\mathbf{z}^{(l)} \in \mathbb{R}^3$ are generally expressed in a transformed version of the physical coordinate system.

\subsection{Augmented Channel Charting}
To enable absolute user localization based on channel charting, we employ the concept of augmented channel charting \cite{asilomar2023}, which incorporates classical localization methods into the channel charting pipeline to learn a channel chart that lies inherently in the physical coordinate frame.
At first, the previously computed dissimilarity matrix is scaled on the basis of the position estimates obtained from triangulation, such that the dissimilarities approximate the corresponding physical distances.
Then, the augmented loss function $\mathcal{L}_\mathrm{aug}$ is defined as a combination of a slightly modified version of the Siamese loss (without normalization) and $\mathcal{L}_\mathrm{AoA}$ as
\begin{align*}
        \mathcal{L}_\mathrm{aug} = \sum_{i, j} \left(1 - \lambda \right) \left(d_{i, j} - \lVert \mathbf z^{(i)} - \mathbf z^{(j)} \rVert \right)^2\\- \lambda \left(\mathcal L_\mathrm{AoA}\left(\mathbf z^{(i)}\right) + \mathcal L_\mathrm{AoA} \left(\mathbf z^{(j)}\right)\right),
\end{align*}
where $\lambda$ is a hyperparameter that weights the individual loss terms.

\subsection{Multistory Channel Charting}
We introduce multistory channel charting, an extension to conventional channel charting, to enhance the performance in multistory buildings.
The extended pipeline, as illustrated in Fig.~\ref{fig:cc_pipeline_multistory}, involves the classification of the floor on which the user is located, followed by applying conventional channel charting separately to each individual floor.

\subsubsection{Floor Classification}
In a multistory building, the channel chart positions $\mathbf{z}^{(l)} \in \mathbb{R}^3$ obtained by conventional channel charting already embed the height of the \ac{UE} to some extent.
We apply the k-means algorithm \cite{macqueen1967multivariate} to this embedding for self-supervised clustering, whereas the number of clusters is given by the number of floors $N_\mathrm{floor}$ in the building.
As a result, we obtain the estimated cluster index $\hat n_\mathrm{cluster}^{(l)} \in \{1,\ldots,N_\mathrm{floor}\}$ for all $l = 1,\ldots,L$.
Note that the cluster indices can be permuted compared to the true floor indices due to the self-supervised nature.
In practice, however, it is often sufficient to determine whether two datapoints have been acquired on the same floor, while the exact floor index is of minor importance.
Otherwise, clusters can be assigned to the actual floors using either a few labeled datapoints or heuristics based on the \acp{AoA} or received powers at the \ac{BS} arrays.

\subsubsection{Forward Charting Function for each Floor}
We partition the dataset into $N_\mathrm{floor}$ subsets, where each subset corresponds to an estimated floor.
For each floor with index $n_\mathrm{floor} \in \{1, \ldots, N_\mathrm{floor}\}$, we define the subset of datapoints as:
\[
    \mathcal{D}_{n_\mathrm{floor}} = \left\{ l \in \{1, \ldots, L\} \mid \hat{n}_\mathrm{cluster}^{(l)} = n_\mathrm{floor} \right\}.
\]
For each floor, a separate \ac{FCF} is learned, which maps the respective \ac{CSI} feature vectors to two-dimensional channel chart positions:
\[
    \mathbf{z}^{(l)} = \mathcal C_{\boldsymbol{\theta}, n_\mathrm{floor}}\left(\mathbf{f}^{(l)}\right) \in \mathbb{R}^2, \quad \forall l \in \mathcal{D}_{n_\mathrm{floor}}.
\]
Optionally, an individual height value can be assigned to each floor to obtain three-dimensional channel chart positions.

\subsection{Novel Beamspace CSI Features}\label{sec:beamspace_features}
We propose a novel feature engineering method designed to extract sparse features from the beamspace \ac{CSI} that are suitable for localization.
At first, we apply zero-padding along the spatial dimensions of the frequency-domain \ac{CSI} $\mathbf{H}^{(l)} \in \mathbb{C}^{B \times M_\mathrm{row} \times M_\mathrm{col} \times N_\mathrm{sub}}$ to double their length, followed by a 2D-\ac{FFT} to transform the spatial dimensions into the beamspace domain:
\[
    \bar{\mathbf{H}}^{(l)} = \mathcal{F}_{\mathrm{2D}} \left( \mathbf{H}_\mathrm{ZP}^{(l)} \right) \in \mathbb{C}^{B \times U_\mathrm{el} \times U_\mathrm{az} \times N_\mathrm{sub}},
\]
where $U_\mathrm{el} = 2 M_\mathrm{row}$ and $U_\mathrm{az} = 2 M_\mathrm{col}$ denote the resulting elevation and azimuth beam dimensions, respectively, and $\mathbf{H}^{(l)}_\mathrm{ZP}$ is the zero-padded version of $\mathbf{H}^{(l)}$.
To extract meaningful information, we compute two separate features.
The mean power per beam $\mathbf{P}^{(l)}$ highlights the direction of dominant paths by aggregating the signal energy over all subcarriers:
\[
    \mathbf{P}_{b,u_\mathrm{el},u_\mathrm{az}}^{(l)} = \sum_{n=1}^{N_\mathrm{sub}} \left\lvert \bar{\mathbf{H}}_{b, u_\mathrm{el}, u_\mathrm{az}, n}^{(l)}\right\rvert^2.
\]
Additionally, we compute a coarse estimate for the time of arrival $\mathbf{D}^{(l)}$ for each angular bin:
\[
    \mathbf{D}_{b,u_\mathrm{el},u_\mathrm{az}}^{(l)} = \arg \left( \sum_{n=1}^{N_\mathrm{sub}-1} \bar{\mathbf{H}}_{b,u_\mathrm{el},u_\mathrm{az},n+1}^{(l)} \left(\bar{\mathbf{H}}_{b,u_\mathrm{el},u_\mathrm{az},n}^{(l)} \right)^* \right).
\]
We concatenate these features to obtain the final feature vector
\[
    \mathbf{f}^{(l)} = \mathrm{vec}\begin{pmatrix} \mathbf{P}^{(l)} \\ \mathbf{D}^{(l)} \end{pmatrix} \in \mathbb{R}^{2 \cdot B \cdot U_\mathrm{el} \cdot U_\mathrm{az}},
\]
which serves as input for the \ac{DNN}.

\begin{figure*}
    \centering
    \begin{subfigure}[b]{0.245\textwidth}
        \centering
        \includegraphics[width=\textwidth, clip]{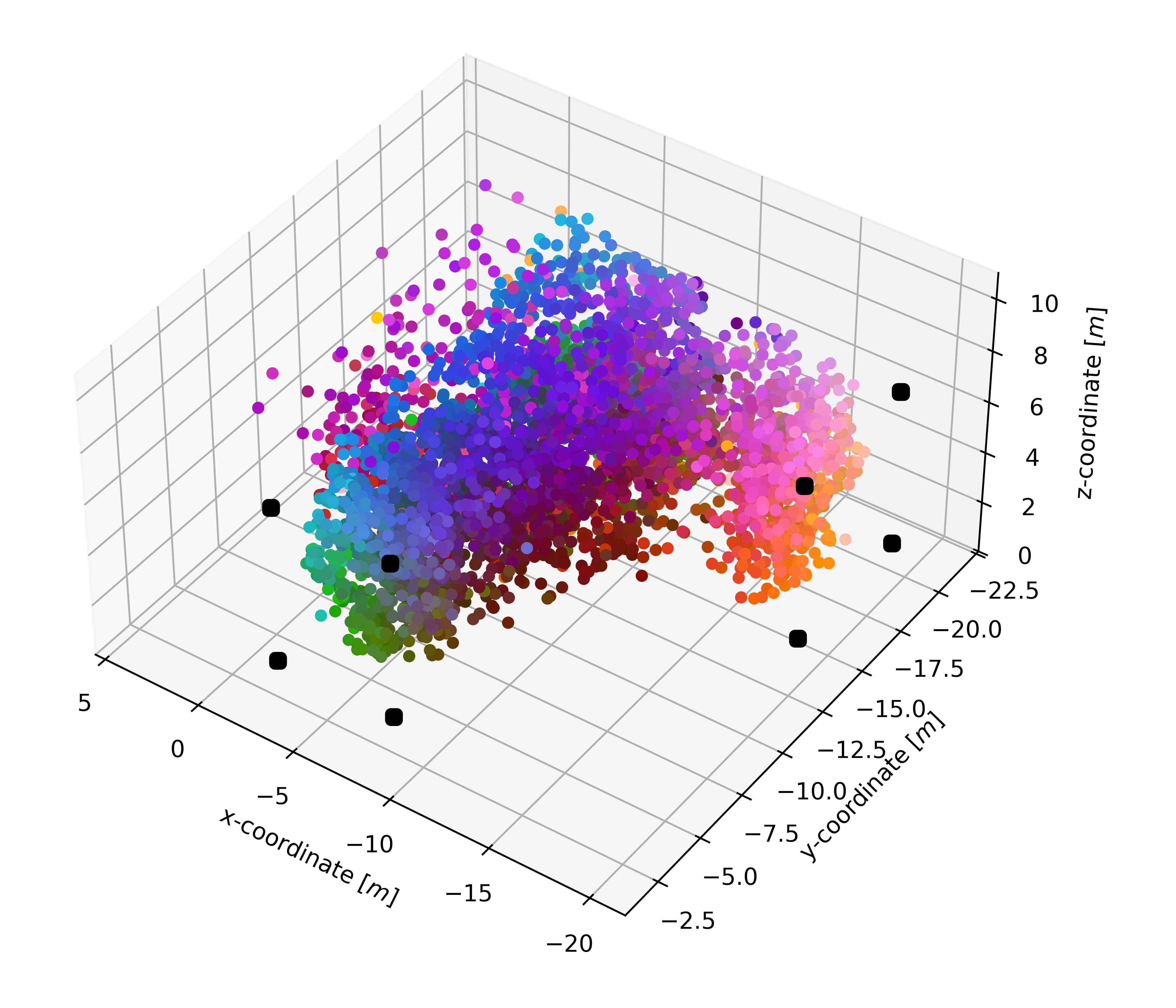}
        \vspace{-0.25cm}
        \caption{Classical localization}
        \label{fig:classical_positions}
    \end{subfigure}
    \begin{subfigure}[b]{0.245\textwidth}
        \centering
        \includegraphics[width=\textwidth, clip]{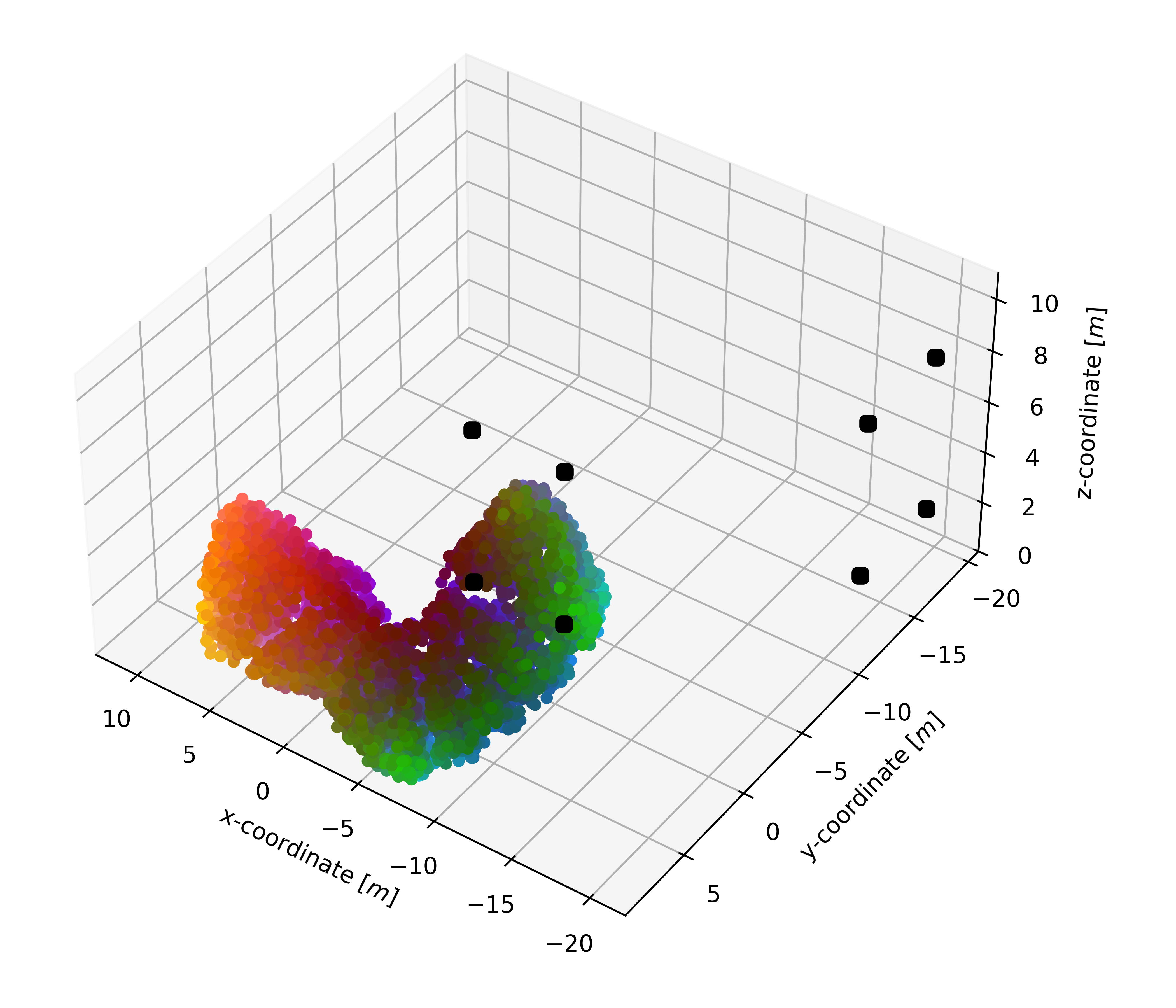}
        \vspace{-0.25cm}
        \caption{Channel charting}
        \label{fig:cc_positions}
    \end{subfigure}
    \begin{subfigure}[b]{0.245\textwidth}
        \centering
        \includegraphics[width=\textwidth, clip]{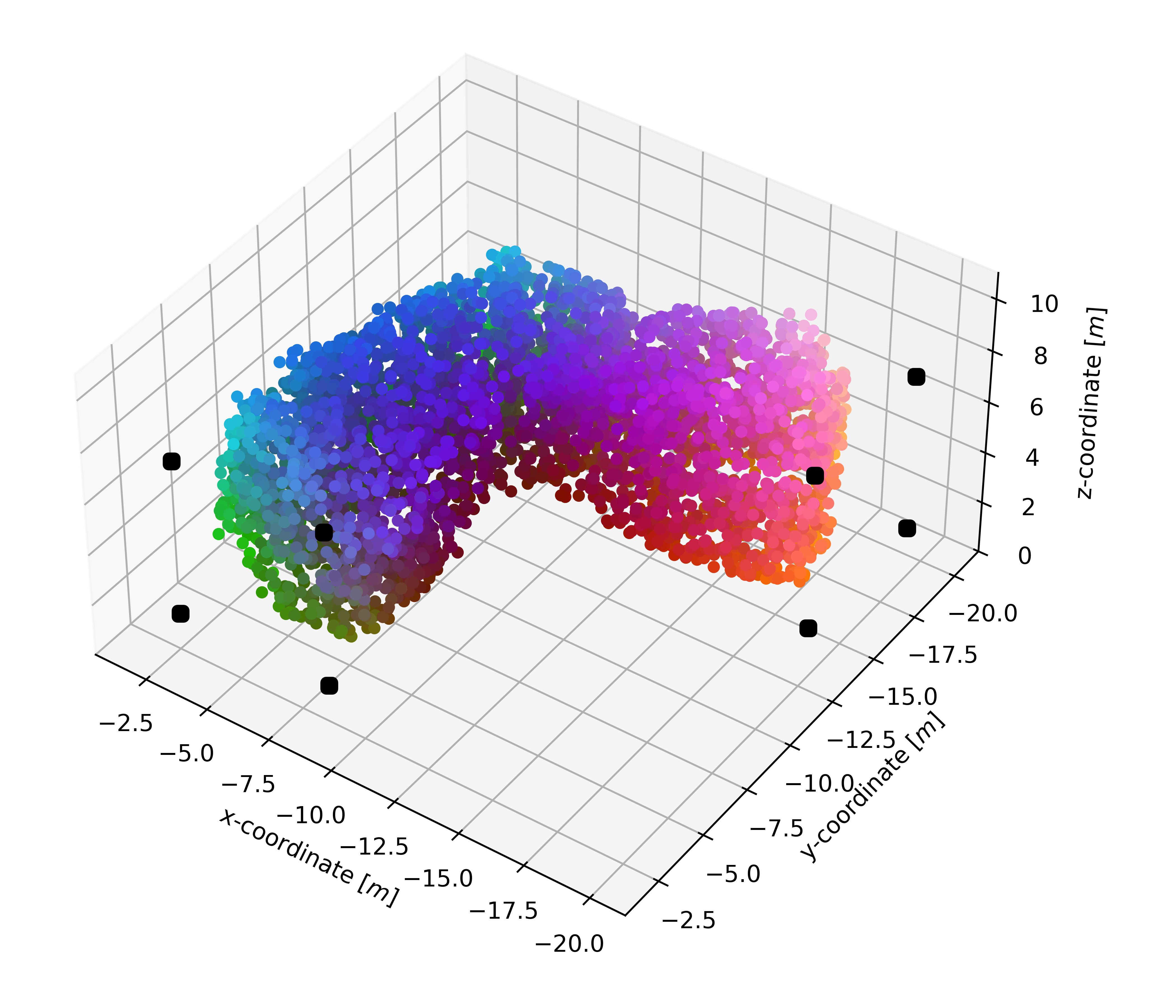}
        \vspace{-0.25cm}
        \caption{Channel charting (transformed)}
        \label{fig:cc_positions_transformed}
    \end{subfigure}
    \begin{subfigure}[b]{0.245\textwidth}
        \centering
        \includegraphics[width=\textwidth, clip]{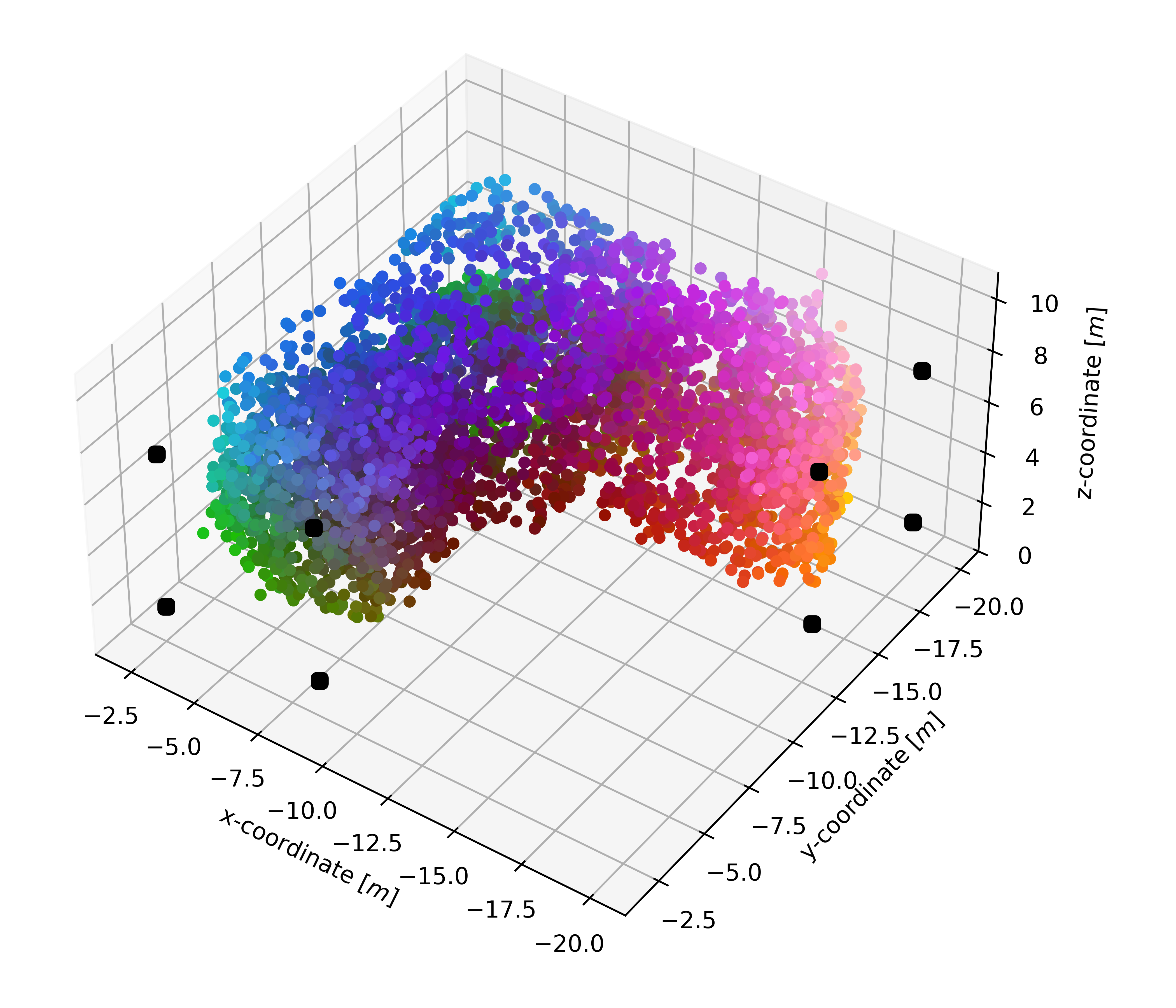}
        \vspace{-0.25cm}
        \caption{Augmented channel charting}
        \label{fig:augmented_cc_positions}
    \end{subfigure}
    \vspace{-0.55cm}
    \caption{Scatter plots of the position estimates for scenario 1 with color gradient preserved from the ground truth positions in Fig.~\ref{fig:dataset_scenario1_positions}: (a) classical localization, (b) channel charting, (c) channel charting followed by an optimal affine tranformation w.r.t. the ground truth positions, and (d) augmented channel charting.}
    \label{fig:estimated_positions_scenario1}
\end{figure*}

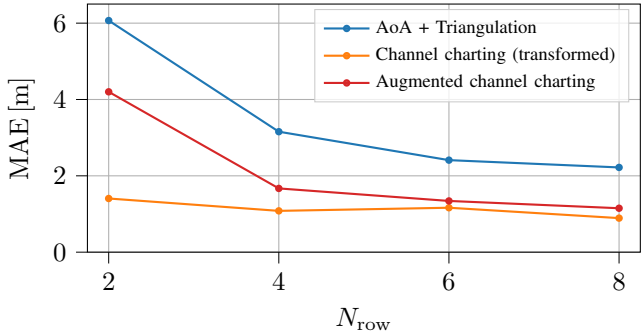
\begin{figure}
    \centering
\begin{tikzpicture}

\definecolor{crimson2143940}{RGB}{214,39,40}
\definecolor{darkgray176}{RGB}{176,176,176}
\definecolor{darkorange25512714}{RGB}{255,127,14}
\definecolor{forestgreen4416044}{RGB}{44,160,44}
\definecolor{lightgray204}{RGB}{204,204,204}
\definecolor{steelblue31119180}{RGB}{31,119,180}

\begin{axis}[
legend cell align={left},
legend style={fill opacity=0.8, draw opacity=1, text opacity=1, draw=lightgray204},
tick align=outside,
width=\columnwidth,height=0.55\columnwidth,
tick pos=left,
x grid style={darkgray176},
xlabel={$N_\mathrm{row}$},
xmajorgrids,
xmin=1.75, xmax=8.25,
xtick style={color=black},
xtick={2,4,6,8},
y grid style={darkgray176},
ylabel={$\mathrm{MAE} \left[\si{\meter}\right]$},
ymajorgrids,
ymin=0.0, ymax=6.5,
ytick style={color=black}
]
\addplot [thick, steelblue31119180, mark=*, mark size=1, mark options={solid}]
table {%
2 6.068
4 3.156
6 2.411
8 2.220
};
\addlegendentry{\scriptsize{AoA + Triangulation}}
\addplot [thick, darkorange25512714, mark=*, mark size=1, mark options={solid}]
table {%
2 1.406
4 1.082
6 1.164
8 0.892
};
\addlegendentry{\scriptsize{Channel charting (transformed)}}
\addplot [thick, crimson2143940, mark=*, mark size=1, mark options={solid}]
table {%
2 4.201
4 1.670
6 1.342
8 1.151
};
\addlegendentry{\scriptsize{Augmented channel charting}}
\end{axis}

\end{tikzpicture}
    \vspace{-0.2cm}
    \caption{Scenario 1: MAE vs. elevation resolution.}
    \label{fig:MAE_elevation_res}
\end{figure}

\begin{figure}
    \centering
\begin{tikzpicture}

\definecolor{crimson2143940}{RGB}{214,39,40}
\definecolor{darkgray176}{RGB}{176,176,176}
\definecolor{darkorange25512714}{RGB}{255,127,14}
\definecolor{forestgreen4416044}{RGB}{44,160,44}
\definecolor{lightgray204}{RGB}{204,204,204}
\definecolor{steelblue31119180}{RGB}{31,119,180}

\begin{axis}[
legend cell align={left},
legend style={fill opacity=0.8, draw opacity=1, text opacity=1, draw=lightgray204},
tick align=outside,
width=\columnwidth,height=0.55\columnwidth,
tick pos=left,
x grid style={darkgray176},
xlabel={Datapoints per $\si{\meter}^3$},
xmajorgrids,
xmin=0.075, xmax=1.025,
xtick style={color=black},
xtick={0.1,0.2,0.3,0.4,0.5,0.6,0.7,0.8,0.9,1.0},
y grid style={darkgray176},
ylabel={$\mathrm{MAE} \left[\si{\meter}\right]$},
ymajorgrids,
ymin=0.75, ymax=4.25,
ytick style={color=black}
]
\addplot [thick, steelblue31119180, mark=*, mark size=1, mark options={solid}]
table {%
0.1 2.344
0.2 2.337
0.5 2.446
1.0 2.439
};
\addlegendentry{\scriptsize{AoA + Triangulation}}
\addplot [thick, darkorange25512714, mark=*, mark size=1, mark options={solid}]
table {%
0.1 3.109
0.2 2.472
0.5 1.390
1.0 1.046
};
\addlegendentry{\scriptsize{Channel charting (transformed)}}
\addplot [thick, crimson2143940, mark=*, mark size=1, mark options={solid}]
table {%
0.1 3.416
0.2 1.896
0.5 1.757
1.0 1.438
};
\addlegendentry{\scriptsize{Augmented channel charting}}
\end{axis}

\end{tikzpicture}
    \vspace{-0.2cm}
    \caption{Scenario 1: MAE vs. datapoint density.}
    \label{fig:MAE_datapoints_3D}
\end{figure}
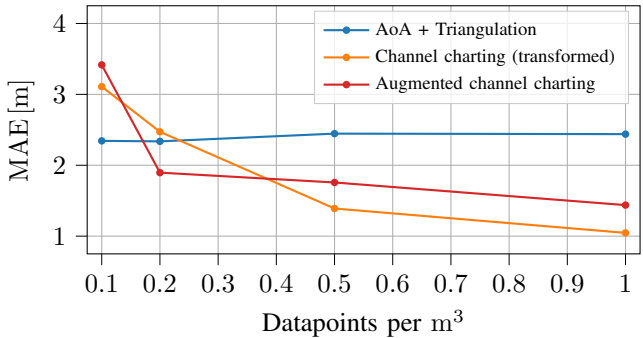

\begin{figure}
    \centering
    \begin{subfigure}[b]{0.49\columnwidth}
        \centering
        \includegraphics[width=\textwidth, clip]{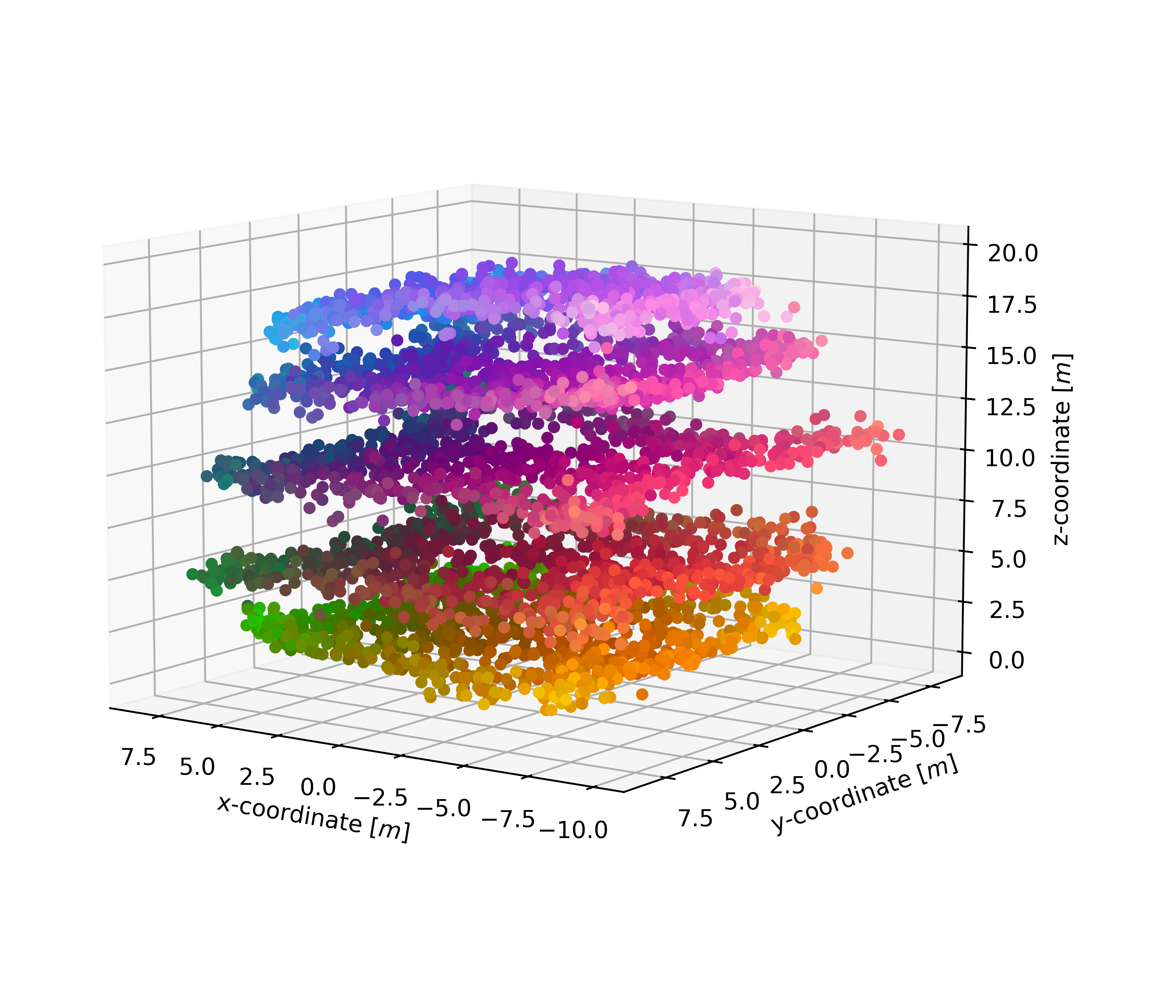}
        \vspace{-0.45cm}
        \caption{Conventional channel charting}
        \label{fig:cc_positions_multistory_transformed_naive}
    \end{subfigure}
    \begin{subfigure}[b]{0.49\columnwidth}
        \centering
        \includegraphics[width=\textwidth, clip]{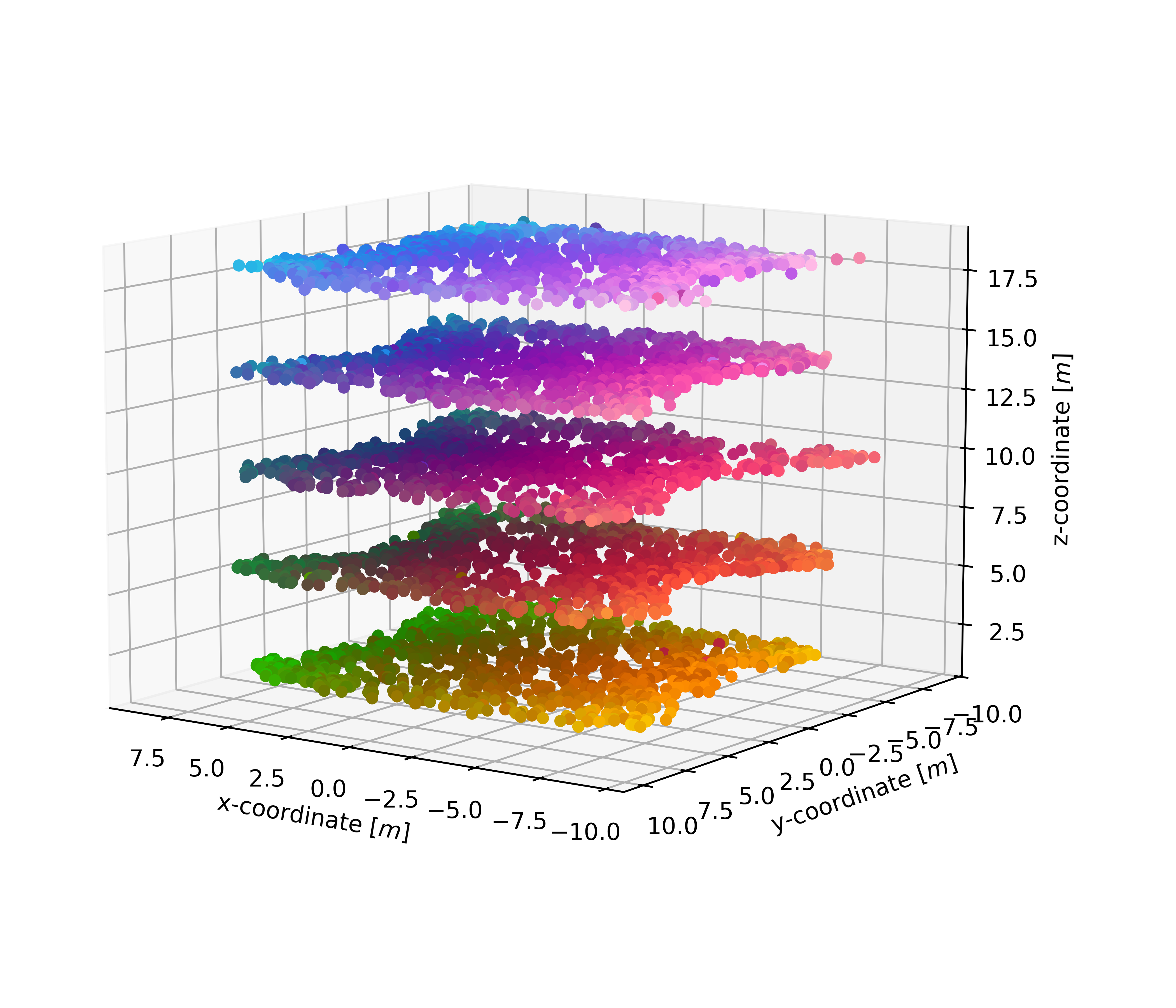}
        \vspace{-0.45cm}
        \caption{Multistory channel charting}
        \label{fig:cc_positions_multistory_transformed_new}
    \end{subfigure}
    \vspace{-0.5cm}
    \caption{Scatter plot of the transformed channel chart positions for scenario 2 with color gradient preserved from the ground truth positions in Fig.~\ref{fig:dataset_scenario2_positions}: (a) conventional channel charting, and (b) multistory channel charting.}
    \label{fig:estimated_positions_scenario2}
\end{figure}

\section{Results}\label{sec:results}
Conventional and augmented channel charting for three-dimensional localization are evaluated in scenario 1 to assess potential challenges in a three-dimensional setting.
In scenario 2, conventional and multistory channel charting are comparatively evaluated in the environment of a multistory building.

\subsection{Evaluation Metric}\label{sec:evaluation_metric}
We use the absolute localization error as a performance metric, which is more meaningful for localization than other common metrics for channel charting.
Since conventional channel charts do not lie in physical coordinates, an optimal affine transformation $T_\mathrm{optimal}(\mathbf z) = \hat { \mathbf A } \mathbf z + \hat { \mathbf b }$ must first be applied to align the channel chart positions $\mathbf z^{(l)}$ with the ground truth positions $\mathbf x^{(l)}$ \cite{fraunhofer_cc}.
To find the optimal transformation parameters $(\mathbf{\hat A}, \mathbf{\hat b})$, we solve the least squares problem
\[
    (\mathbf{\hat A}, \mathbf{\hat b}) = \argmin\limits_{(\mathbf{A}, \mathbf{b})} \sum_{l = 1}^L \lVert\mathbf{A} \mathbf z^{(l)} + \mathbf b - \mathbf x^{(l)} \rVert^2.
\]
The \ac{MAE} is then computed as the mean Euclidean distance between the transformed channel chart positions and the ground truth positions:
\[
    \mathrm{MAE} = \frac{1}{L} \sum_{l=1}^L \lVert T_\mathrm{optimal}(\mathbf z^{(l)}) - \mathbf x ^{(l)} \rVert.
\]
Note that the \ac{MAE} for position estimates obtained by triangulation and augmented channel charting is computed without prior affine transformation.

\subsection{Scenario 1: Three-dimensional Localization}
Fig.~\ref{fig:estimated_positions_scenario1} shows scatter plots of the position estimates for the different localization methods in scenario 1.
Classical triangulation, as depicted in Fig.~\ref{fig:classical_positions}, yields rough position estimates achieving a \ac{MAE} of $2.26\, \si{\meter}$.
Conventional channel charting produces position estimates that are not embedded in physical coordinates (Fig.~\ref{fig:cc_positions}).
Applying the optimal affine transformation to the channel chart positions, as described in Section~\ref{sec:evaluation_metric}, they can be interpreted in physical coordinates, as shown in Fig.~\ref{fig:cc_positions_transformed}.
The \ac{MAE} of $0.90\, \si{\meter}$ is significantly lower compared to triangulation.
Augmented channel charting (Fig.~\ref{fig:augmented_cc_positions}) learns positions directly in physical coordinates and achieves a \ac{MAE} of $1.17\, \si{\meter}$.
The performance is comparable to the transformed conventional channel chart positions and significantly better than the triangulation results, highlighting the advantage of combining classical localization with channel charting in a three-dimensional environment.

\subsubsection{Impact of Angular Resolution}
Fig.~\ref{fig:MAE_elevation_res} shows the \ac{MAE} for the compared methods as a function of the number of antenna rows $N_\mathrm{row}$ per array, which determine the angular resolution in elevation direction.
As expected, the \ac{MAE} for triangulation increases significantly with fewer antenna rows.
Channel charting, on the other hand, is barely affected, showing its ability to compensate for low angular resolution in elevation through sheer datapoint density.
Augmented channel charting is slightly affected for $N_\mathrm{row} \geq 4$, but suffers substantially for $N_\mathrm{row} = 2$, similar to triangulation.
This can be attributed to the reliance of augmented channel charting on accurate triangulation estimates to learn positions within physical coordinates.

\subsubsection{Impact of Datapoint Density}
As shown in Fig.~\ref{fig:MAE_datapoints_3D}, the performance of both conventional and augmented channel charting is highly dependent on the datapoint density, showing a significant decrease for lower datapoint densities, whereas classical triangulation stays unaffected.
This presents a practical limitation of channel charting, as a sufficiently high datapoint density across the entire physical space of interest may not be guaranteed.
Although channel charting faces the same challenge if applied in two dimensions, achieving a consistently high datapoint density is substantially more difficult in three-dimensional environments.

\subsection{Scenario 2: Multistory Channel Charting}
Fig.~\ref{fig:estimated_positions_scenario2} depicts the estimated channel chart positions after an optimal affine transformation for both conventional and multistory channel charting.
The conventional channel charting method already produces estimates that roughly separate the different floors, resulting in a \ac{MAE} of $1.42\, \si{\meter}$.
These position estimates are used by multistory channel charting for floor classification, yielding a classification error rate of $0.42\%$.
Despite this fraction of datapoints that are assigned to an incorrect floor, the overall localization accuracy is substantially improved, with a \ac{MAE} of $0.99\, \si{\meter}$.

\section{Conclusion and Outlook}\label{sec:conclusion}
This work applies the concept of augmented channel charting for three-dimensional localization and demonstrates that channel charting can compensate for limited angular resolution at the \ac{BS} and that the effectiveness of channel charting depends on a sufficiently high datapoint density.
The proposed multistory channel charting approach significantly improves localization accuracy in multistory buildings.
Furthermore, we have introduced a new feature engineering method for localization based on beamspace \ac{CSI}.
Although the results on ray tracing-based \ac{CSI} are promising, the proposed concepts should be validated on real-world channel measurements.
Moreover, the antenna configurations and radio environments considered in this work are particularly favorable for localization, so future work should investigate alternative antenna setups and a wider range of environments.
Where available, incorporating timestamp information could further improve performance in both considered scenarios.
Another open research question is the analysis of how antenna polarization affects channel charting in both two-dimensional and three-dimensional settings.

\bibliographystyle{IEEEtran}
\bibliography{IEEEabrv,references}

\begin{thebibliography}{10}
\providecommand{\url}[1]{#1}
\csname url@samestyle\endcsname
\providecommand{\newblock}{\relax}
\providecommand{\bibinfo}[2]{#2}
\providecommand{\BIBentrySTDinterwordspacing}{\spaceskip=0pt\relax}
\providecommand{\BIBentryALTinterwordstretchfactor}{4}
\providecommand{\BIBentryALTinterwordspacing}{\spaceskip=\fontdimen2\font plus
\BIBentryALTinterwordstretchfactor\fontdimen3\font minus \fontdimen4\font\relax}
\providecommand{\BIBforeignlanguage}[2]{{%
\expandafter\ifx\csname l@#1\endcsname\relax
\typeout{** WARNING: IEEEtran.bst: No hyphenation pattern has been}%
\typeout{** loaded for the language `#1'. Using the pattern for}%
\typeout{** the default language instead.}%
\else
\language=\csname l@#1\endcsname
\fi
#2}}
\providecommand{\BIBdecl}{\relax}
\BIBdecl

\bibitem{wen_5g_localization_survey}
F.~Wen, H.~Wymeersch, B.~Peng, W.~P. Tay, H.~C. So, and D.~Yang, ``{A Survey on 5G Massive MIMO Localization},'' \emph{Digital Signal Processing}, vol.~94, pp. 21--28, 2019.

\bibitem{savic2015fingerprinting}
V.~Savic and E.~G. Larsson, ``{Fingerprinting-Based Positioning in Distributed Massive MIMO Systems},'' in \emph{2015 IEEE 82nd Vehicular Technology Conference (VTC2015-Fall)}, 2015, pp. 1--5.

\bibitem{vieira2017deep}
J.~Vieira, E.~Leitinger, M.~Sarajlic, X.~Li, and F.~Tufvesson, ``{Deep Convolutional Neural Networks for Massive MIMO Fingerprint-Based Positioning},'' in \emph{2017 IEEE 28th Annual International Symposium on Personal, Indoor, and Mobile Radio Communications (PIMRC)}, 2017.

\bibitem{cc_features_ferrand}
P.~Ferrand, A.~Decurninge, and M.~Guillaud, ``{DNN-based Localization from Channel Estimates: Feature Design and Experimental Results},'' in \emph{2020 IEEE Global Communications Conference}, 2020, pp. 1--6.

\bibitem{studer_cc}
C.~Studer, S.~Medjkouh, E.~G{\"o}n{\"u}lta{\c{s}}, T.~Goldstein, and O.~Tirkkonen, ``{Channel Charting: Locating Users Within the Radio Environment Using Channel State Information},'' \emph{IEEE Access}, vol.~6, pp. 47\,682--47\,698, 2018.

\bibitem{kazemi_cc_snr_prediction}
P.~Kazemi, H.~Al-Tous, C.~Studer, and O.~Tirkkonen, ``{SNR Prediction in Cellular Systems based on Channel Charting},'' in \emph{2020 IEEE Eighth International Conference on Communications and Networking (ComNet)}, 2020, pp. 1--8.

\bibitem{yassine_beam_prediction}
T.~Yassine, B.~Chatelier, V.~Corlay, M.~Crussière, S.~Paquelet, O.~Tirkkonen, and L.~Le~Magoarou, ``{Model-Based Deep Learning for Beam Prediction Based on a Channel Chart},'' in \emph{2023 57th Asilomar Conference on Signals, Systems, and Computers}, 2023, pp. 1636--1640.

\bibitem{wcnc2025}
P.~Stephan, F.~Euchner, and S.~ten Brink, ``{Channel Charting-Based Channel Prediction on Real-World Distributed Massive MIMO CSI},'' in \emph{2025 IEEE Wireless Communications and Networking Conference (WCNC)}, 2025, pp. 1--6.

\bibitem{pihlajasalo2020absolute}
J.~Pihlajasalo, M.~Koivisto, J.~Talvitie, S.~Ali-L{\"o}ytty, and M.~Valkama, ``{Absolute Positioning with Unsupervised Multipoint Channel Charting for 5G Networks},'' in \emph{2020 IEEE 92nd Vehicular Technology Conference (VTC2020-Fall)}.\hskip 1em plus 0.5em minus 0.4em\relax IEEE, 2020, pp. 1--5.

\bibitem{taner_cc_real_world_coordinates}
S.~Taner, V.~Palhares, and C.~Studer, ``{Channel Charting in Real-World Coordinates With Distributed MIMO},'' \emph{IEEE Transactions on Wireless Communications}, vol.~24, no.~9, pp. 7286--7300, 2025.

\bibitem{asilomar2023}
F.~Euchner, P.~Stephan, and S.~ten Brink, ``{Augmenting Channel Charting with Classical Wireless Source Localization Techniques},'' in \emph{57th Asilomar Conference on Signals, Systems, and Computers}, 2023.

\bibitem{karmanov2021wiclusterpassiveindoor2d3d}
I.~Karmanov, F.~G. Zanjani, I.~Kadampot, S.~Merlin, and D.~Dijkman, ``{WiCluster: Passive Indoor 2D/3D Positioning using WiFi without Precise Labels},'' in \emph{2021 IEEE Global Communications Conference (GLOBECOM)}, 2021, pp. 1--7.

\bibitem{sionna}
J.~Hoydis, S.~Cammerer, F.~{Ait Aoudia}, A.~Vem, N.~Binder, G.~Marcus, and A.~Keller, ``{Sionna: An Open-Source Library for Next-Generation Physical Layer Research},'' \emph{arXiv preprint}, Mar. 2022.

\bibitem{dataset-dichasus-adxx}
\BIBentryALTinterwordspacing
F.~Euchner, P.~Stephan, M.~Gauger, and S.~ten Brink, ``{CSI Dataset dichasus-adxx: ARENA2036: Distributed setup in industrial environment at 3.4GHz},'' 2024. [Online]. Available: \url{https://doi.org/doi:10.18419/darus-4062}
\BIBentrySTDinterwordspacing

\bibitem{henninger_3D_localization}
M.~Henninger, T.~E. Abrudan, S.~Mandelli, M.~Arnold, S.~Saur, V.-M. Kolmonen, S.~Klein, T.~Schlitter, and S.~ten Brink, ``{Probabilistic 5G Indoor Positioning Proof of Concept with Outlier Rejection},'' in \emph{2022 Joint European Conference on Networks and Communications \& 6G Summit (EuCNC/6G Summit)}, 2022, pp. 249--254.

\bibitem{stephan2024angle}
P.~Stephan, F.~Euchner, and S.~ten Brink, ``{Angle-Delay Profile-Based and Timestamp-Aided Dissimilarity Metrics for Channel Charting},'' \emph{IEEE Transactions on Communications}, vol.~72, no.~9, pp. 5611--5625, 2024.

\bibitem{macqueen1967multivariate}
J.~MacQueen, ``{Some Methods for Classification and Analysis of Multivariate Observations},'' in \emph{Proceedings of the 5th Berkeley Symposium on Mathematical Statistics and Probability}, vol.~1, 1967, pp. 281--297.

\bibitem{fraunhofer_cc}
M.~Stahlke, G.~Yammine, T.~Feigl, B.~M. Eskofier, and C.~Mutschler, ``{Indoor Localization With Robust Global Channel Charting: A Time-Distance-Based Approach},'' \emph{IEEE Transactions on Machine Learning in Communications and Networking}, vol.~1, pp. 3--17, 2023.

\end{thebibliography}

\end{document}